\documentclass[doublecol]{epl2} 
\usepackage{graphicx}
\usepackage{dcolumn}
\usepackage{bm}
\usepackage{amssymb,amsmath,latexsym,amsfonts}%
\title{{Planck Scale Physics and Bogoliubov Spaces in a Bose--Einstein Condensate}}
\shorttitle{Planck Scale Physics and Bogoliubov Spaces in a Bose--Einstein Condensate} 

\author{E. Castellanos\inst{1}}
\shortauthor{E. Castellanos}

\institute{Departamento de
F\'{\i}sica, Centro de Investigaci\'on y Estudios Avanzados del IPN\\
 A. P. 14--740,  07000, M\'exico, D.F., M\'exico.                  
  \inst{1}\email{ecastellanos@fis.cinvestav.mx} }

\abstract{We analyze the consequences caused by a deformed dispersion relation, suggested in several quantum gravity models, upon a bosonic gas. Concerning the ground state of the Bogoliubov space of this system, we deduce the corrections in the pressure, the speed of sound, and the corresponding healing length. Indeed, we prove that the corrections in the relevant thermodynamic properties associated with the ground state, defines a non trivial function of the density of particles and the deformation parameters, allowing us to constrain, in principle, the form of the modified energy--momentum dispersion relation.}

\begin{document}

\maketitle

\section{Introduction}

The possibility of a deformation in the dispersion relation of microscopic particles, appears in connection with the quest for a quantum theory of gravity \cite{ami,Giovanni1,Claus,Claus1,Kostelecky,amelino1,5,12,13}. In some schemes, the possibility that the space-time could be quantized, can be
characterized, from a phenomenological point of view, as a
modification in the dispersion relation of microscopic particles
\cite{Giovanni1,Kostelecky,Claus,nico,nico1,nico2} (and references therein). A modified dispersion relation
emerges as an adequate tool in the search for phenomenological
consequences caused by this type of quantum gravity models.
Nevertheless, the most difficult aspect in searching experimental hints relevant for the quantum-gravity problem is the
smallness of the involved effects \cite{Kostelecky,amelino1}. If this kind of deformations are
characterized by some Planck scale, then the quantum gravity effects
become very small \cite{Giovanni1,Claus}. 

In the non--relativistic
limit, the deformed dispersion relation can be expressed as follows
\cite{Claus,Claus1}
\begin{equation}
E \simeq
m+\frac{p^{2}}{2m}+\frac{1}{2M_{p}}\Bigl(\xi_{1}mp+\xi_{2}p^{2}+\xi_{3}\frac{p^{3}}{m}\Bigr),
\label{ddr}
\end{equation}
 in units where the speed of light $c=1$, being $M_{p}$ ($\simeq 1.2\times 10^{28}eV$) the Planck mass.
The three parameters $\xi_{1}$, $\xi_{2}$, and $\xi_{3}$, are model
dependent \cite{Giovanni1,Claus}, and should take\, positive or
negative values close to $1$. There are some evidence within the
formalism of Loop quantum gravity \cite{Claus,Claus1,5,12} that
indicates a non--zero values for the three parameters, $\xi_{1},\,
\xi_{2},\, \xi_{3}$, and particulary \cite{5,13} that produces a
linear--momentum term in the non--relativistic limit. Unfortunately,
as is usual in quantum gravity phenomeno-\,logy, the possible bounds
associated with the deformation parameters, open a wide range of
possible magnitudes, which is translated to a significant challenge. 

In a previous report \cite{Castellanos}, we were able to prove, that the condensation temperature associated with the non--interacting bosonic gas, trapped in a generic three dimensional power law potential, is corrected as a consequence of the deformation in the dispersion relation. Moreover, this correction described as a non-trivial function of the number of particles and the shape associated to the corresponding trap it could provide representative bounds for the deformation parameter $\xi_{1}$. We have proved in reference \cite{Castellanos} (see also \cite{r1,r2}) that the deformation parameter $\xi_{1}$ can be bounded, under typical conditions, from $|\xi_{1}|\lesssim10^{6}$ for a linear trap, to $|\xi_{1}|\lesssim10^{2}$ corresponding to a bosonic gas in a box. In the case of a harmonic oscillator--type potential, we have obtained a bound up to  $|\xi_{1}|\lesssim10^{4}$.

In references \cite{Claus,Claus1} it was suggested
the use of ultra-precise cold-atom-recoil experiments to constrain
the form of the energy-momentum dispersion relation in the
non--relativistic limit. There, the bound associated to $\xi_{1}$ is at least, four orders of magnitude smaller
than the bound associated to a Bose--Einstein condensate trapped in a harmonic oscillator obtained in reference \cite{Castellanos}. However, the results obtained in reference \cite{Castellanos}, suggest that the many body contributions would allow to improve, in principle, the bounds associated with $\xi_{1}$. These facts open the possibility to explore the corrections caused by an anomalous dispersion relation, on the relevant thermodynamic properties associated with N--body systems, for instance, within the Bogoliubov formalism associated with Bose--Einstein condensates.

At this point we must add that many body systems as a tool in the search of quantum gravity manifestations, through modifications of the uncertainty principle, has been analyzed in the context of the center of mass motion of macroscopic bodies \cite{I,JD,AM}. 
Nevertheless, in reference \cite{AM}, it is was argued that the approach followed in references \cite{I,JD}, cannot be used in this context, due to an \emph{incorrect extrapolation criterion} of Planck--scale space--time quantization for fundamental particles, to macroscopic bodies. As a consequence, the author in reference \cite{AM} suggests that the center of mass motion of a macroscopic body should be affected more weakly than its constituents due to the $N^{-s}$ ($N$ is the number of particles in the macroscopic system, and $s$ is some positive power) suppression of the corresponding corrections caused by the quantum structure of space time. Indeed, as the author in reference \cite{AM} remarks, this is not a generic criterion.

Here it is important to emphasize that the main difference between the approach followed in the present manuscript and the one followed in references \cite{I,JD,AM}, lies in the fact that the corrections caused by a deformed dispersion relation is a collective behavior of all the particles forming the condensate, nor the properties of a single point, like the center of mass motion, for example. In other words, the corrections caused by the deformation parameters are analyzed over some properties of the entire system (the condensate itself) and in some cases, like the ground state energy or the corresponding speed of sound (see below), these corrections scale as a non--trivial function of the number of particles. Additionally, the approach followed in the present manuscript suggests that this type of macroscopic systems, could be more sensitive in some cases, to Planck scale manifestations than its constituents.

On the other hand, the phenomenon of Bose--Einstein condensation, from
the theoretical and experimental point of view, has produced an
enormous amount of relevant and interesting publications associated to this topic, see for instance 
\cite{Dalfovo,Andrews,Andrews2,RP,elias1,Pitaevski,Pethick,Ueda,Griffin}, and references therein. Among the issues addressed we may find its possible use as tools in gravitational physics or in searching quantum--gravity
manifestations, for instance, in the context of Lorentz violation or
to provide phenomenological constrains on Planck--scale physics
\cite{Colladay,Donald,cam1,elca,Camacho,CastellanosCamacho,CastellanosCamacho1,CastellanosClaus,Castellanos,r1,r2,eli}.

For this purpose, we define the following \emph{modified}
N-body Hamiltonian
\begin{eqnarray}
\hat{H} =
\sum_{\vec{k}=0}\Big\{\frac{\hbar^2k^2}{2m}
+\alpha\hbar k+mc^{2}\Bigl\}\,\, \hat{a}_{\vec{k}}^{\dagger}\hat{a}_{\vec{k}}\nonumber\\
+\frac{U_0}{2V}\sum_{\vec{k}=0}\sum_{\vec{p}=0}\sum_{\vec{q}=0}\hat{a}_{\vec{p}}^{\dagger}\hat{a}_{\vec{q}}^{\dagger}
 \hat{a}_{\vec{p}+\vec{k}}\hat{a}_{\vec{q}-\vec{k}} , \label{Ham1}
\end{eqnarray}
where the creation and annihilation operators satisfy the usual canonical commutation relations for bosons
\begin{equation}
[\hat{a}_{\vec{k}},\hat{a}_{\vec{k}'}^{\dagger}]=\delta_{\vec{k}\vec{k}'},\,\,\,\ [\hat{a}_{\vec{k}},\hat{a}_{\vec{k}'}]=[\hat{a}_{\vec{k}}^{\dagger},\hat{a}_{\vec{k}'}^{\dagger}]=0.
\end{equation}
Additionally, $U_0=\frac{4\pi \hbar^2}{m}a$, being $a$ the s--wave scattering length, $mc^{2}$ the rest energy, and  the term $\alpha \hbar k$ with $\alpha=\xi_{1}\frac{mc}{2M_{p}}$, in ordinary units, is the leading order modification in expression (\ref{ddr}). Clearly, if we set $\alpha=0$, we recover the usual expression \cite{Pethick,Ueda}.
The system under study will be a Bose--Einstein condensate
enclosed in a container of volume $V$, having a deformed dispersion relation as a single particle energy spectrum. Moreover, besides low temperature, to use the s-scattering approximation, it is also required a dilute gas system, i.e., $n|a|^{3} << 1$, where $a$ is $s$--scattering lenght \cite{Pitaevski}.

The main goal of this work is to analyze the corrections caused by a deformed dispersion relation upon the properties associated with the ground state of a Bose--Einstein condensate, within the Bogoliubov formalism, i.e., the corrections in the ground state energy, and consequently the corrections in the associated speed of sound. In other words, we will show that the many--body contributions associated to this system, could be used, in principle, to explore possible Planck scale manifestations on some relevant properties associated with the condensate. In addition, we will show that the present approach opens the possibility to explore alternative scenarios compared to those suggested in references \cite{CastellanosClaus,Castellanos,r1,r2,I,JD,AM}.

\section{Modified Bogoliubov Spaces}

Let us calculate the corrections on some relevant thermodynamical properties, caused by the deformation parameter $\alpha=\xi_{1}\frac{mc}{2M_{p}}$ in a Bose--Einstein condensate within the Bogolioubov formalism  (see for instance, \cite{Pathria,Ueda}). Let us assume that most of the particles are in the condensate, that is, in the $\vec{k}=0$ state, or equivalently, the number of particles in the excited states is negligible. These last assertions can be expressed as follows
\begin{equation}
N_{0}\approx N, \,\,\,\,\,\,\,\, \sum_{\vec{k} \not=0} N_{\vec{k}} <<N,
\end{equation}
being $N$ the total number of particles, $N_{\vec{k}}$ the number of particles in the excited states, and $N_{0}$ the number of particles in the ground state. Keeping terms up to second order in $\hat{a}_{0}$, the Hamiltonian (\ref{Ham1}) becomes
\begin{eqnarray}
\hat{H} &=&\frac{U_0N^2}{2V}+mc^{2}N
\nonumber\\&+&\sum_{\vec{k}\not=0}\Bigl[\frac{\hbar^2k^2}{2m}+\alpha\hbar k+mc^{2}+\frac{U_0N}{V}\Bigr]\hat{a}_{\vec{k}}^{\dagger}\hat{a}_{\vec{k}}\nonumber\\&+&2\hat{a^{\dag}_{0}} \hat{a}_{0} \sum_{\vec{k}\not=0}\hat{a}_{\vec{k}}^{\dagger}\hat{a}_{\vec{k}}+\hat{a_{0}^{\dag^{2}}}\sum_{\vec{k}\not=0}\hat{a}_{\vec{k}}\hat{a}_{-\vec{k}}
+ \hat{a_{0}}^{2} \sum_{\vec{k}\not=0 }\hat{a}_{\vec{k}}^{\dagger}\hat{a}_{-\vec{k}}^{\dagger}.\,\,\,\,\,\,\,\,
\label{Add2}
\end{eqnarray}
Finally, in the same order of the approximation we assume that $\hat{a}_{0}^{\dag} \hat{a}_{0},\ \hat{a}_{0}^{2},\ 
 \hat{a_{0}^{\dag^{2}}} = N$. Using these facts, the \emph{modified} Hamiltonian (\ref{Ham1}) can be re--expressed as follows
\begin{eqnarray}
\hat{H} &= &\frac{U_0N^2}{2V}+mc^{2}N
\nonumber\\&+&\sum_{\vec{k}\not=0}\Bigl[\frac{\hbar^2k^2}{2m}+\alpha\hbar k+mc^{2}+\frac{U_0N}{V}\Bigr]\hat{a}_{\vec{k}}
^{\dagger}\hat{a}_{\vec{k}}\nonumber\\
&+&\sum_{\vec{k}\not=0}\frac{U_0N}{2V}\Bigl[\hat{a}_{\vec{k}}^{\dagger}\hat{a}_{-\vec{k}}^{\dagger}
+ \hat{a}_{\vec{k}}\hat{a}_{-\vec{k}}\Bigr]. \label{Ham2}
\end{eqnarray}
In order to  obtain the ground state energy associated with our system, let us diagonalize the above \emph{modified} Hamiltonian by introducing  the so--called Bogoliubov
transformations \cite{Pethick,Pitaevski,Ueda}
\begin{equation}
\hat{a}_{\vec{k}}=
\frac{\hat{b}_{\vec{k}} -
\gamma_k\hat{b}_{-\vec{k}}^{\dagger}}{\sqrt{1-\gamma_k^2}},\label{Bog1} \hspace{0.5cm}\hat{a}_{\vec{k}}^{\dagger}=
\frac{\hat{b}_{\vec{k}}^{\dagger} -
\gamma_k\hat{b}_{-\vec{k}}}{\sqrt{1-\gamma_k^2}}.
\end{equation}
The operators $\hat{b}_{\vec{k}}^{\dagger}$ and $\hat{b}_{\vec{k}}$ are the creation and annihilation operators associated with a quasiparticle called \emph{Bogolon} \cite{Ueda} and it is straightforward to show that these operators also obey the canonical commutation relations for bosons as $\hat{a}_{\vec{k}}^{\dagger}$ and $\hat{a}_{\vec{k}}$ \cite{Pathria}. Inserting the Bogoliubov
transformations (\ref{Bog1}) into equation (\ref{Ham2}), we are able to obtain, after some algebra, the following diagonalized Hamiltonian
\begin{eqnarray}
\hat{H} = \frac{U_0N^2}{2V}+mc^{2}N+ \sum_{\vec{k}\not=0}\sqrt{\epsilon_{k_{\alpha}}\Bigl(\epsilon_{k_{\alpha}}+\frac{2U_0N}{V}\Bigl)}
\hat{b}_{\vec{k}}^{\dagger}\hat{b}_{\vec{k}}\nonumber\\
+\sum_{\vec{k}\not=0}\Bigg\{-\frac{1}{2}\Bigg[\frac{U_0N}{V} +\epsilon_{k_{\alpha}}
-\sqrt{\epsilon_{k_{\alpha}}\Bigl(\epsilon_{k_{\alpha}}+\frac{2U_0N}{V}\Bigr)}\,\,\Bigg]\Bigg\}, \,\,\label{Ham3}
\end{eqnarray}
where $\epsilon_{k_{\alpha}} =\frac{\hbar^2k^2}{2m}+\alpha\hbar k+mc^{2}$. Notice that in our case the \emph{modified} Bogoliubov coefficient $\gamma_{k}$  is given by
\begin{equation}
\gamma_k = 1+ \frac{V\epsilon_{k_{\alpha}}}{U_0N}
-\sqrt{\frac{V\epsilon_{k_{\alpha}}}{U_0N}}\sqrt{2+
\frac{V\epsilon_{k_{\alpha}}}{U_0N}}. \label{Add33}
\end{equation}
In the usual case, $\alpha=0$, the last summation in the Hamiltonian (\ref{Ham3}) diverges as $(U_{0}N/V)^{2}/2\epsilon_{k}$\cite{Gribakin,Ueda}, as can be seen by performing an expansion of the last term in equation (\ref{Ham3}) for large $k$. When $\alpha=0$, we have that $\epsilon_{k}=\frac{\hbar^{2}k^{2}}{2m}$, and the divergence disappears by introducing the so--called pseudo--potential method \cite{Ueda}. 
 The pseudo--potential $U_{P}(\bold{r})$ can be expressed as follows

 \begin{equation}
 \label{PP}
 U_{P}(\bold{r})=U_{0}\delta(\bold{r})\frac{\partial}{\partial r}r,
 \end{equation} 
being $\delta(\bold{r})$ the Dirac Delta function. When $\alpha=0$, the leading terms in the divergence are of order $1/k^{2}$, which corresponds to a $1/r$ behavior in the configuration space. Thus, by using the pseudo--potential (\ref{PP}) applied to the $1/r$ behavior, shows that divergence can be removed.
 
Notice that in our case,  the divergence is of order $1/((1/2m)\tilde{k}^{2}+mc^2-m\alpha^{2}/2)$, where we have defined $\tilde{k}=k+m\alpha$. Thus, for all practical purposes, the divergence can be expressed as $\sim 1/\tilde{k}^{2}$ without loss of generality. 

The 3-D Fourier transform in spherical coordinates can be written as \cite{MC,Grad}
\begin{equation}
V(r)=\Bigl(\frac{2}{\pi}\Bigr)^{1/2} \int_{0}^{\infty} V(k) \frac{k \sin(kr) }{r} \,dk.
\end{equation}
Thus, the Fourier transform upon $1/ \tilde{k}^{2}$ can be expressed generically to first order in $\alpha$ as follows 
\begin{eqnarray}
\label{FT}
V(r) &=& \Bigl(\frac{2}{\pi}\Bigr)^{1/2} \int_{m\alpha}^{\infty} \frac{(\tilde{k}-m\alpha) \sin \Bigl((\tilde{k}-m\alpha)r\Bigr) }{r \tilde{k}^{2}} \,d\tilde{k} \nonumber\\ &\sim& \frac{1}{r}+\frac{f(r,\alpha)}{r}m\alpha+O(\alpha^{2}),
\end{eqnarray}
where $f(r,\alpha)$ is given by
\begin{equation}
f(r,\alpha)=\frac{\pi}{2}+2(\gamma+ \ln r+\ln (m\alpha))r-r,
\end{equation}
being $\gamma$ the Euler's constant \cite{Grad}. Here it is noteworthy to mention that even in this situation the action of the operator (\ref{PP}), removes the divergence in the Fourier transform (\ref{FT}). In fact, $f(r,\alpha)\rightarrow 0$ when $r\rightarrow 0$. Additionally, when $\alpha \rightarrow 0$, we recover the usual behavior $1/r$.

Thus, by subtracting the divergent term $(U_{0}N/V)^{2}/2\epsilon_{k_{\alpha}}$ in the corresponding Hamiltonian (\ref{Ham3}), we safely remove the divergent behavior. Using these facts, our \emph{modified} Hamiltonian now is given by
\begin{eqnarray}
\label{Ham301}
\hat{H}& =& \frac{U_0N^2}{2V}+mc^{2}N+ \sum_{\vec{k}\not=0}\sqrt{\epsilon_{k_{\alpha}} \Bigl(\epsilon_{k_{\alpha}}+\frac{2U_0N}{V}\Bigl)} 
\hat{b}_{\vec{k}}^{\dagger}\hat{b}_{\vec{k}}\,\,\,\,\,\,\,\,\,\,\,\,
\\\nonumber&+&\sum_{\vec{k}\not=0}\Bigg\{-\frac{1}{2}\Bigg[\frac{U_0N}{V} +\epsilon_{k_{\alpha}}
-\sqrt{\epsilon_{k_{\alpha}}\Bigl(\epsilon_{k_{\alpha}}+\frac{2U_0N}{V}\Bigr)}\\\nonumber&-& \Bigl(\frac{U_0N}{ V}\Bigr)^2\frac{1}{2 \epsilon_{k_{\alpha}}}\Bigg]\Bigg\}. 
\end{eqnarray}
Replacing the last summation in the above equation by an integration, using the standard prescription, $\sum_{k} \rightarrow V \int d \vec{k}/(2\pi \hbar)^{3}$ \cite{Pethick}, and introducing a dimensionless variable $x$ defined by
\begin{equation}
x = \sqrt{\frac{\epsilon_{k_{\alpha}}V}{U_0N}},\label{Psedo33}
\end{equation}
 leads us to the following expression associated with the ground state energy of the system   
\begin{eqnarray}
E_{0}&= &\frac{U_0N^2}{2V}+mc^{2}N-\Bigg\{\frac{2\pi V (2m)^{3/2}}{(2\pi\hbar)^{3}}\Bigl(\frac{U_{0}N}{V}\Bigr)^{5/2}\nonumber\\&\times&\int_{\delta}^{\infty}x^{2}\Bigl[1+x^2-x\sqrt{2+x^2}-\frac{1}{2x^2}\Bigr]dx \Bigg\}\nonumber\\&+&
\alpha \Bigg\{\frac{8 \pi m^{2}V }{(2\pi\hbar)^{3}}\Bigl(\frac{U_{0}N}{V}\Bigr)^{2}\nonumber\\&\times&\int_{\delta}^{\infty}x\Bigl[1+x^2-x\sqrt{2+x^2}-\frac{1}{2x^2}\Bigr]dx\Bigg\}\nonumber\\&+&O(\delta^{2},\alpha^{2}),
\label{Psedo2}
\end{eqnarray}
where
\begin{eqnarray}
\delta =\sqrt{mc^{2}\Bigl(\frac{V}{U_{0}N}\Bigr)}.\label{Psedo33}
\end{eqnarray}
Notice that if we set $\alpha=0$, then we recover the usual result \cite{Ueda,Pathria}, corrected by the contributions of the rest energy.  
Finally, from equation (\ref{Psedo2}) we are able to obtain the energy of the ground state
\begin{eqnarray}
E_0 &=&\frac{2\pi a\hbar^2N^2}{mV}\Bigg[1 +
8\sqrt{2}\sqrt{\frac{a^3N}{V\pi}}\Bigg(\frac{8\sqrt{2}}{15}-\frac{1}{2}\delta\Bigg)\\\nonumber&
+&\alpha\frac{2am}{\pi^{2} \hbar}\Bigl(0.05+\ln \delta \Big)\Bigg]+mc^{2}N+O(\delta^{2},\alpha^{2}).
\label{E0}
\end{eqnarray}
Hence, the associated speed of sound reads
\begin{eqnarray}
\label{Sp0} 
v^2_s & =&-\frac{{V}^{2}}{N m}\frac{\partial P_{0}}{\partial
V}=\frac{4\pi a\hbar^2N}{m^2V}\Bigg[1 +
\frac{242}{15}\sqrt{\frac{a^3N}{V\pi}}\\\nonumber&\times& \Bigl(1-\frac{15}{16\sqrt{2}}\delta\Bigr)+\alpha\frac{2am}{\hbar \pi^{2}}\Bigl(\ln \delta-\frac{31}{20}\Bigr)\Bigg]+ O(\delta^{2},\alpha^2),
\end{eqnarray}
where $P_{0}=-\frac{\partial E_{0}}{\partial V}$ is the pressure of the ground state. Let us remark that the contributions of the deformation parameter on the ground state energy and the corresponding speed of sound scales as a non--trivial function of the number of particles together with the deformation parameter $\alpha$. In other words, these results suggest that this type of macroscopic systems, could be more sensitive to Planck scale manifestations than its constituents.

The speed of sound in a condensate is typically of order $10^{-3}$m$s^{-1}$ \cite{Andrews,Andrews2}. This fact allows us to estimate the sensibility to detect Planck scale manifestations for our system when $\vert\xi_{1}\vert\lesssim1$ under typical conditions, i.e., densities of order $10^{13}-10^{15}$ atoms per $cm^{3}$ \cite{Dalfovo,Griffin}, $a \sim 10^{-9}$m \cite{Dalfovo}, and $m=15 \times 10^{-26}$ kg, in the case of $^{39}_{19} K$. In this situation the corresponding speed of sound associated with the deformation parameter could be inferred up to $10^{-10}$ $m\,s^{-1}$. 
Conversely, densities of order $10^{18}$ atoms per $cm^{3}$, are needed in order to obtain a typical speed of sound of order $10^{-3}$$m\,s^{-1}$, that is, three orders of magnitude bigger than typical densities, when $\vert\xi_{1}\vert\lesssim1$.  Notice also that, the correction upon the speed of sound caused by $\alpha$, is a non--trivial function of the density $n$, which is proportional to the product $\alpha\,  n \ln n^{-1/2}$, this relation between the deformation parameter $\alpha$ and the density $n$, suggests that denser systems are needed in order to make our system more sensitive to Plank scale signals when  $|\xi_{1}| \lesssim 1$.

On the other hand, from the \emph{modified }$N$--body Hamiltonian (\ref{Ham301}), we are able to recognize the \emph{modified}--energy of the Bogoliubov excitations
\begin{equation}
\label{Ek}
E_{k} = \sqrt{\epsilon_{k_{\alpha}}\Bigl(\epsilon_{k_{\alpha}}+
 \frac{2U_0N}{V}\Bigr)}. 
\end{equation}
The long--wavelength limit, $k\rightarrow 0$, associated with the above expression, leads to the following phonon--like dispersion relation corrected by the contributions of the deformation parameter $\alpha$
\begin{equation}
\label{qpart}
E_{k_{\alpha}}=\Bigl[\frac{\hbar^{2}k}{2m}+\frac{\alpha \hbar}{2}\Bigr]\sqrt{\frac{16\pi a N}{V}},
\end{equation}
Conversely, at the high--energy limit, $k\rightarrow \infty$,  expression (\ref{Ek}) reads
\begin{equation}
\label{part}
E_{k_{\alpha}}=\frac{\hbar^{2}k^{2}}{2m}+\alpha\hbar k+\frac{4\pi \hbar^{2}N a}{mV},
\end{equation}
which basically is the deformed dispersion relation, expression (\ref{ddr}), corrected by the mean field contributions. If we set $\alpha=0$ in equations (\ref{qpart}) and (\ref{part}), we recover the usual result \cite{Ueda}.

The crossover between the phonon spectrum (\ref{qpart}) and the single particle behavior (\ref{part}) related to the \emph{modified}--energy of the Bogoliubov excitations (\ref{Ek}),   
allows to define the corresponding healing length
\begin{equation}
\xi_{\alpha} \approx \frac{1}{\sqrt{8\pi an}}\Bigl(1- \alpha \frac{m}{\hbar \sqrt{8\pi a n}}\Bigr).
\label{HL0}
\end{equation}
Under typical conditions, the healing length is approximately (or bigger than) one micrometer \cite{pana}. The order of magnitude associated with the correction caused by the deformation term is of order $10^{-5} \,m$ for $\vert \xi_{1}\vert\lesssim 1$ in the case of $^{39}_{19} K$ for typical densities $10^{13}-10^{15}$ atoms per $cm^{3}$. Notice that more diluted systems could be used to improve the correction caused by the deformation parameter in this context. If we set $\alpha=0$ in equation (\ref{HL0}) then,  we recover the usual expression \cite{Pathria,Ueda,Pitaevski,Pethick}.

\section{Conclusions}

We have analyzed a modified bosonic gas in a box within the Bogoliubov formalism. We have calculated the corrections on the ground state energy of this system, and consequently the corrections in the associated speed of sound. The corresponding expression associated with the speed of sound (\ref{Sp0}) suggests that denser systems would allow to improve the sensibility to detect Planck scale manifestations caused by the quantum structure of space--time, but where the diluteness condition, $n |a| ^{3} << 1$, remains valid. In addition, the corrections on the healing length can be estimated up to $10^{-5}$m, under typical conditions assuming $\vert \xi_{1} \vert \lesssim 1$, which is notable.

Clearly, the results given in this work, must be extended to a more realistic situation. First, from the experimental point of view, there is no condensates in a box. Usually, the confinement of the condensate can be obtained by using harmonic traps, among others \cite{Dalfovo,Pethick} and the use of generic potentials, under certain conditions, could be used to constrain Planck scale physics in this context, by changing the shape of the trap \cite{CastellanosClaus,Castellanos}. In other words, the analysis of a system trapped in generic potential within the Bogoliubov formalism deserves further investigation, and will be presented elsewhere \cite{elias2}. 
Finally, we must add that the measurement of the corrections caused by the deformation parameter $\alpha$, could be out of the current technology. Nevertheless, it is remarkable that the many--body contributions in a Bose--Einstein condensate, open the possibility of planning specific scenarios that could be used, in principle, to test a possible manifestation of the effects caused by the quantum structure of space--time in low energy earth based experiments.

\acknowledgments
This work was partially supported by CONACyT M\'exico under grants
CB-2009-01, no. 132400, CB-2011, no. 166212,  and I0101/131/07
C-234/07 of the Instituto Avanzado de Cosmolog\'ia (IAC)
collaboration (http://www.iac.edu.mx/). E. C. acknowledges CONACyT for the grant received.

\end{document}